\documentclass[aps,prb,10pt,twocolumn,showpacs]{revtex4-1}

\usepackage{graphicx}

\begin{document}

\title{Crystal structure and properties of barium thorate BaThO$_3$ from
first principles}

\author{Alexander I. Lebedev}
\email[]{swan@scon155.phys.msu.ru}
\affiliation{Physics Department, Moscow State University,
Leninskie gory, 119991 Moscow, Russia}

\date{\today}

\begin{abstract}
The phonon spectrum of cubic BaThO$_3$ with the perovskite structure is
calculated from first principles within the density functional theory. The
analysis of unstable modes in the phonon spectrum enables to determine the
symmetry of all possible distorted phases, calculate their energies, and
show that the ground-state structure of barium thorate is $Pbnm$. For this
structure, the static and optical dielectric constants, elastic moduli,
heat capacity, Raman spectra, and the energy band gap in the LDA and $GW$
approximations are calculated. The possibility of the structural phase
transitions in BaThO$_3$ is also discussed.
\end{abstract}


\maketitle

\section{Introduction}

The $AB$O$_3$ perovskite oxides constitute an important family of materials that
display interesting physical properties (ferroelectricity, ferroelasticity,
superconductivity, magnetism) and are widely used in technological applications.
Barium thorate BaThO$_3$ is one of the members of this family.
High thermal stability of barium thorate (its melting point is above
2300$^\circ$C) and relatively low work function of thorium enable to use BaThO$_3$
for thermionic cathodes of high-intensity discharge lamps (see, for example,
patents~\cite{USPatent.2394095, USPatent.3029359, USPatent.3188236}). Barium thorate
ceramics doped with neodymium~\cite{SolidStateIonics.70.291, SolidStateIonics.178.461}
and yttrium~\cite{SolidStateIonics.79.183} are solid electrolytes, which have
the highest proton conductivity at temperatures above 900$^\circ$C among solid
electrolytes. Since barium thorate is produced as the fission product in nuclear
reactors (especially in uranium--thorium reactors), the knowledge of its properties
is also important for predicting the reliability of nuclear fuel elements.

Insufficient knowledge of the properties of BaThO$_3$ becomes evident from
the fact that even its crystal structure is not well established. In the
very first work,~\cite{ZPhysChemB.28.65} the structure of barium thorate was
identified as a simple perovskite structure with the lattice parameter of
$a \approx 4.48$~{\AA}. A more detailed study~\cite{ActaCryst.13.653} found
the superstructure reflections on the diffraction patterns, but because of the
lack of splitting of the strong peaks, the structure was considered as a
pseudocubic one with a doubled lattice parameter. Finally, after the splitting
of the main reflections was observed,~\cite{ChemLett.3.429, JNuclMater.280.51}
the structure of BaThO$_3$ was identified as orthorhombic, but neither the
space group nor the atomic coordinates were determined.

Thermodynamic properties of barium thorate (the Gibbs energy of formation)
were determined in Refs.~\onlinecite{JNuclMater.275.201, JAlloysComp.290.97}.
In Ref.~\onlinecite{OptMater.33.553},  the electronic structure, optical and
elastic properties of BaThO$_3$ were calculated from first principles using
the FP-LAPW approach, but for some reason the calculations were limited to
the cubic five-atom unit cell, whereas X-ray studies clearly indicate the
lower symmetry. According to these calculations, the cubic barium thorate
is a direct-gap insulator with the band gap of $E_g = 5.7$~eV. This result,
however, contradicts the density-of-states calculations presented in
Ref.~\onlinecite{OptMater.33.553},
according to which the energy gap between the valence and conduction bands is
3.33~eV, and the calculations of the $\epsilon_2(\omega)$ optical spectra in
which the absorption starts at an energy below 5~eV.

The knowledge of the true crystal structure of BaThO$_3$ is crucial for correct
prediction of its properties. In this work, we use the first-principles
calculations to determine the equilibrium structure of barium thorate, calculate
some of its properties, and discuss the possibility of the structural phase
transitions in it.

\section{Calculation details}

To predict the properties of BaThO$_3$, we must first determine its ground-state
structure which has the lowest energy at $T = 0$. For this purpose, we must
first calculate the phonon spectrum of its parent $Pm3m$ phase and then, by
adding the distortions corresponding to unstable modes in the phonon spectrum
to the structure, seek for a minimum-energy structure in which the energy
of all optical phonons at all points of the Brillouin zone are positive and the
structure is mechanically stable (the stability criterion is the positive values
of the determinant and all leading principal minors of the 6$\times$6 matrix
of elastic moduli in the Voigt notation).~\cite{PhysSolidState.51.362}

\begin{table*}
\caption{\label{table1}Electronic configuration of atoms and parameters used
for construction of pseudopotentials: $r_s$, $r_p$, and $r_d$ are radii of
the pseudopotential cores for $s$-, $p$-, and $d$-projections; $q_s$, $q_p$,
and $q_d$ are the cut-off wave vectors used in the optimization procedure;
$r_{\rm min}$, $r_{\rm max}$, and $V_{\rm loc}$ are the range limits and depth
of the correcting local potential. All the values are in Hartree atomic units
except for the correction energy which is in Ry.}
\begin{ruledtabular}
\begin{tabular}{ccccccccccc}
Atom & Configuration      & $r_s$ & $r_p$ & $r_d$ & $q_s$ & $q_p$ & $q_d$ & $r_{\rm min}$ & $r_{\rm max}$ & $V_{\rm loc}$ \\
\hline
Ba   & $5s^25p^65d^06s^0$ & 1.85  & 1.78  & 1.83  & 7.07  & 7.07  & 7.07  & 0.1  & 1.68 & 1.95 \\
Th   & $6s^26p^66d^07s^0$ & 1.88  & 2.04  & 2.04  & 7.57  & 7.27  & 7.07  & 0.01 & 1.76 & 0.75 \\
O    & $2s^22p^43d^0$     & 1.40  & 1.55  & 1.40  & 7.07  & 7.57  & 7.07  & ---  & ---  & --- \\
\end{tabular}
\end{ruledtabular}
\end{table*}

In this work, the first-principles calculations were performed within the
density functional theory (DFT) using the \texttt{ABINIT} software.~\cite{abinit3}
The exchange-correlation interaction was described in the local density approximation
(LDA). Pseudopotentials for Ba and O atoms used in the calculations were taken
from Ref.~\onlinecite{PhysSolidState.51.362}. Scalar-relativistic pseudopotential
for the Th atom was constructed using the RRKJ scheme~\cite{PhysRevB.41.1227}
with the \texttt{OPIUM} program.~\cite{opium}  To improve the transferability of
the pseudopotential, the $s$ local potential correction~\cite{PhysRevB.59.12471}
was used. The parameters used for the
construction of pseudopotentials are presented in Table~\ref{table1}. Testing of
the Th pseudopotential using ThO$_2$ as an example revealed its high enough quality:
the calculated lattice parameter of this compound (5.606~{\AA}) differed from the
experimental value by only 0.15\% and the bulk modulus (205~GPa) differed by 4\%.

The lattice parameters and the equilibrium atomic coordinates in the unit cell
were determined from the condition that the residual forces acting on the atoms
are less than $5 \cdot 10^{-6}$~Ha/Bohr (0.25~meV/{\AA}) and the total energy
is calculated self-consistently to an accuracy of better than $10^{-10}$~Ha. The
integration over the Brillouin zone was performed on the 8$\times$8$\times$8
Monkhorst--Pack mesh for the cubic phase or on the meshes with equivalent $k$-point
density for low-symmetry phases. The maximum plane-wave energy was 30~Ha.

The quasiparticle band gap of BaThO$_3$ was calculated using the so-called
one-shot $GW$ approximation.~\cite{RevModPhys.74.601,PhysStatSolidiB.246.1877}
The Kohn--Sham wave functions and energies calculated within DFT-LDA were used
as a zeroth-order approximation. The dielectric matrix
$\epsilon_{\mathbf{GG'}}(\mathbf{q},\omega)$ was computed for a
6$\times$6$\times$6 $\mathbf{q}$-point mesh from the independent-particle
polarizability matrix $P^0_{\mathbf{GG'}}(\mathbf{q},\omega)$ calculated for
3743~reciprocal-lattice vectors $\mathbf{G}(\mathbf{G'})$, 20~occupied and
280~unoccupied bands. The dynamic screening was described using the Godby--Needs
plasmon-pole model. The components of wave functions with kinetic energy up to
24~Ha were used in these calculations. The energy correction to the DFT-LDA
solution was computed as diagonal matrix elements of the $[\Sigma - E_{xc}]$
operator, where $\Sigma = GW$ is the self-energy operator, $E_{xc}$ is the
exchange-correlation energy operator, $G$ is the Green's function, and
$W = \epsilon^{-1}v$ is the screened Coulomb interaction. In the calculations
of $\Sigma$, the components of wave functions with kinetic energy up to 24~Ha
for both exchange and correlation parts of $\Sigma$ were used. The accuracy of
the band gap calculation estimated from the convergence tests is $\sim$0.05~eV.

\section{Results}

\begin{figure}
\centering
\includegraphics{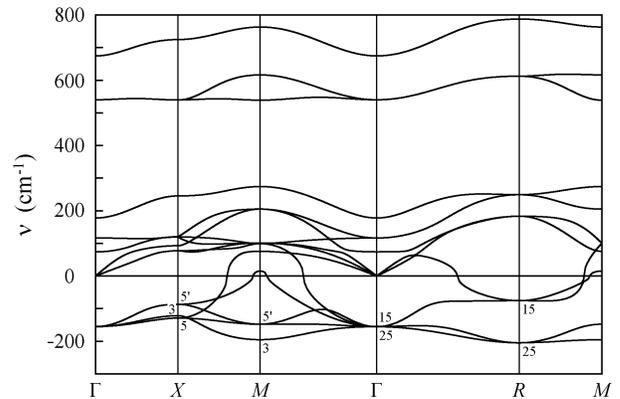}
\caption{Phonon spectrum of BaThO$_3$ in the cubic $Pm3m$ phase. The labels
indicate the symmetry of unstable modes.}
\label{fig1}
\end{figure}

The phonon spectrum of BaThO$_3$ in the cubic $Pm3m$ phase is shown in
Fig.~\ref{fig1}. It is seen that there are nine unstable modes in the phonon
spectrum of barium thorate. They include the triply degenerate $\Gamma_{15}$
mode associated with the ferroelectric instability; the triply degenerate $\Gamma_{25}$
mode describing the uniform deformation of the oxygen octahedra; the triply
degenerate $R_{25}$ mode and nondegenerate $M_3$ mode, which describe the
rotation of the octahedra; the doubly degenerate antiferroelectric $X_5$,
$X'_5$, and $M'_5$ modes; the non-degenerate $X_3$ mode describing spatially
non-uniform deformation of the octahedra; and the triply degenerate $R_{15}$
mode describing the vibrations of the Ba atom in its oxygen environment.

\begin{table}
\caption{\label{table2}The relative energies of the low-symmetry phases of
BaThO$_3$ resulting from the condensation of unstable modes. The most stable
phase is in boldface.}
\begin{ruledtabular}
\begin{tabular}{ccc}
Phase         &	Unstable mode &	Energy (meV) \\
\hline
$Pm3m$        & ---           & 0 \\
$P4_2/mmc$    & $X_3$         & $-$55.9 \\
$Cmcm$        & $X_5'$        & $-$90.9 \\
$Pmma$        & $X_5$         & $-$99.3 \\
$Pmma$        & $X_5'$        & $-$101.4 \\
$Cmcm$        & $X_5$         & $-$118.8 \\
$I4/mmm$      & $R_{15}$      & $-$123.0 \\
$R{\bar 3}m$  & $R_{15}$      & $-$148.0 \\
$P{\bar 4}m2$ & $\Gamma_{25}$ & $-$153.1 \\
$Amm2$        & $\Gamma_{15}; \Gamma_{25}$ & $-$228.0 \\
$Pmma$        & $M_5'$        & $-$234.5 \\
$R32$         & $\Gamma_{25}$ & $-$255.9 \\
$Cmmm$        & $M_5'$        & $-$303.2 \\
$R3m$         & $\Gamma_{15}$ &	$-$326.4 \\
$P4mm$        & $\Gamma_{15}$ &	$-$337.2 \\
$P4/mbm$      & $M_3$         & $-$529.2 \\
$I4/mcm$      & $R_{25}$      & $-$599.3 \\
$Cmcm$        & $R_{25}+M_3$  & $-$696.4 \\
$R{\bar 3}c$  & $R_{25}$      & $-$710.1 \\
$Imma$        & $R_{25}$; $R_{15}$ & $-$754.4 \\
$Pbnm$        & $R_{25}+M_3$  & \bf{$-$797.8} \\
\end{tabular}
\end{ruledtabular}
\end{table}

The energies of all phases obtained from the condensation of the unstable
phonons as well as the energies of $Pbnm$ and $Cmcm$ phases resulting from
simultaneous condensation of the $R_{25}$ and $M_3$ modes are given in
Table~\ref{table2}. Among these phases, the $Pbnm$ phase has the lowest energy.

\begin{table}
\caption{\label{table3}Calculated atomic coordinates in the orthorhombic $Pbnm$
phase of barium thorate.}
\begin{ruledtabular}
\begin{tabular}{ccccc}
Atom & Position & $x$        & $y$     & $z$ \\
\hline
Ba   & $4c$     & $-$0.01246 & 0.03717 & 0.25000 \\
Th   & $4b$     & 0.50000    & 0.00000 & 0.00000 \\
O1   & $4c$     & 0.10447    & 0.46199 & 0.25000 \\
O2   & $4d$     & 0.70713    & 0.29227 & 0.05723 \\
\end{tabular}
\end{ruledtabular}
\end{table}

The calculation of the phonon spectrum in the $Pbnm$ phase shows that the
frequencies of all optical phonons at the center of the Brillouin zone and at
the high-symmetry points on its boundary are positive; the determinant and all
leading principal minors constructed from the elastic tensor are also positive.
This means that the $Pbnm$ phase is the ground-state structure of BaThO$_3$.
The calculated lattice parameters of this phase are $a = 6.3140$, $b = 6.4039$,
and $c = 8.9578$~{\AA}; the atomic coordinates are given in Table~\ref{table3}.
The obtained lattice parameters are in good agreement with the experimental
data of Ref.~\onlinecite{ChemLett.3.429} ($a = 6.345 \pm 0.002$,
$b = 6.376 \pm 0.002 $, $c = 8.992 \pm 0.002$~{\AA}) and
Ref.~\onlinecite{JNuclMater.280.51} ($a = 6.35 \pm 0.01$, $b = 6.387 \pm 0.009$,
$c = 8.995 \pm 0.009$~{\AA}).

We now consider some physical properties of BaThO$_3$ in the ground state.
Although the ferroelectric instability was found in the parent cubic phase, the
structural distortions suppress this instability, and in the ground state barium
thorate is a paraelectric. The static dielectric tensor in the $Pbnm$ phase is
characterized by three diagonal components of $\epsilon^0_{xx} = 24.7$,
$\epsilon^0_{yy} = 25.0$, and $\epsilon^0_{zz} = 37.6$; the tensor of the
optical dielectric constant is described by the components
of $\epsilon^{\infty}_{xx} = 4.42$, $\epsilon^{\infty}_{yy} = 4.35$, and
$\epsilon^{\infty}_{zz} = 4.28$. The elastic moduli are $C_{11} = 195$~GPa,
$C_{22} = 200$~GPa, $C_{33} = 199$~GPa, $C_{12} = 99$~GPa, $C_{13} = 83$~GPa,
$C_{23} = 71$~GPa, $C_{44} = 55$~GPa, $C_{55} = 51$~GPa, $C_{66} = 64$~GPa.
The bulk modulus in the orthorhombic phase is $B = 121.9$~GPa; in the $Pm3m$
phase its value is 124.2~GPa and is very close to the value of 124~GPa
obtained in Ref.~\onlinecite{OptMater.33.553}.

\begin{table}
\caption{\label{table4}Calculated frequencies of the Raman-active modes for
the $Pbnm$ phase of barium thorate.}
\begin{ruledtabular}
\begin{tabular}{cc}
The mode symmetry & Frequency (cm$^{-1}$) \\
\hline
$A_g$    & 75; 89; 157; 207; 300; 341; 523 \\
$B_{1g}$ & 95; 109; 169; 283; 312; 387; 666 \\
$B_{2g}$ & 92; 217; 331; 508; 681 \\
$B_{3g}$ & 96; 123; 298; 520; 613 \\
\end{tabular}
\end{ruledtabular}
\end{table}

\begin{figure}
\centering
\includegraphics{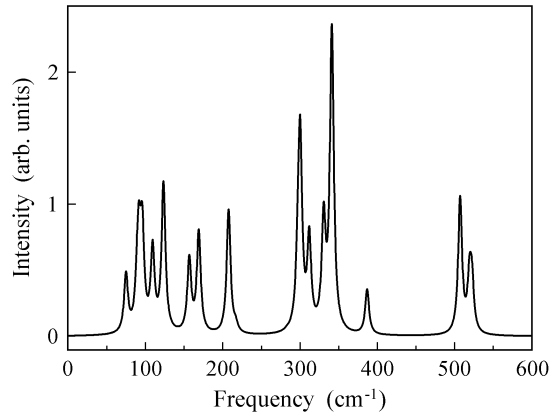}
\caption{Calculated Raman spectra for a polycrystalline sample of BaThO$_3$
at 300~K.}
\label{fig2}
\end{figure}

For identification of the $Pbnm$ phase of BaThO$_3$ it is convenient to use
the Raman spectroscopy. The calculated frequencies of the Raman-active modes
in this phase are given in Table~\ref{table4}, and the theoretical Raman
spectrum calculated for a polycrystalline sample at 300~K using the formulas
from Ref.~\onlinecite{PhysRevB.71.214307} is shown in Fig.~\ref{fig2}.

\begin{figure}
\centering
\includegraphics{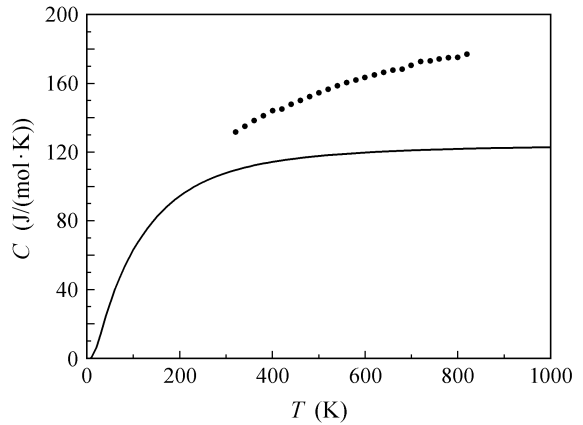}
\caption{Comparison of the calculated heat capacity $C_v$ in the $Pbnm$
phase of barium thorate (solid line) with the experimental
data~\cite{JNuclMater.299.28} (points).}
\label{fig3}
\end{figure}

The energies of different phases obtained in this work (Table~\ref{table2})
enable to make some remarks on the possible phase transitions in barium thorate.
The $Imma$ phase closest in energy to the $Pbnm$ ground-state structure has the
energy only by 43.4~meV higher than the ground-state energy. This means that
a second-order phase transition $Pbnm \to Imma$ can occur in BaThO$_3$ at a
temperature slightly above 300~K. In the experimental study of the heat
capacity,~\cite{JNuclMater.299.28}  no anomalies were detected in the
temperature range 320--820~K, but the peculiarity of the results of
Ref.~\onlinecite{JNuclMater.299.28} is that the
measured values of the specific heat are much higher than our calculated ones
(Fig.~\ref{fig3}) and, more significantly, they even exceed the theoretical
limit of 3$R$ per atom in the unit cell (the Dulong--Petit law). This may
indicate that the investigated sample was likely not pure.%
    \footnote{It is known that, when exposed to air, barium thorate absorbs moisture
    and reacts with CO$_2$ thus forming a mixture of ThO$_2$ + BaCO$_3$.~\cite{ZPhysChemB.28.65,
    ARF-6046-Report}  The annealing temperature before the specific heat
    measurements in Ref.~\onlinecite{JNuclMater.299.28} (550$^\circ$C) was sufficient
    to remove moisture, but was not high enough to decompose BaCO$_3$. After
    the interaction with CO$_2$, the number of atoms in a mole of the substance
    increases, and this results in an increase of the heat capacity.}
We believe that the predicted phase transition was hidden by this fact. In this
regard, we note that the features in the dielectric properties of BaThO$_3$ at
about 260$^\circ$C, which were observed in Ref.~\onlinecite{ARF-6046-Report} and
interpreted as an evidence for the ferroelectric phase transition, can actually
be associated with the $Pbnm \to Imma$ phase transition under discussion.

To resolve the contradictions mentioned in the Introduction on the energy band
gap $E_g$ in cubic BaThO$_3$, we performed our own
calculations of $E_g$ in both the LDA approximation used in this work and
in the $GW$ approximation~\cite{RevModPhys.74.601,PhysStatSolidiB.246.1877}
which takes into account the many-body effects and enables to obtain the energy
band gap in much better agreement with the experiment.~\cite{PhysSolidState.54.1663}
According to our data, in the LDA approximation $E_g^{\rm LDA} = 2.907$~eV in
the $Pm3m$ phase; in the orthorhombic $Pbnm$ phase the band gap increases to
3.947~eV. As the DFT calculations always underestimate the energy band gap,
more realistic $E_g$ values can be obtained in the $GW$ approximation.
These calculations gave the energy band gap $E_g^{GW} \approx 4.40$~eV for
the cubic phase. This value is intermediate between the two values, 5.7~eV and
3.33~eV, reported for this phase in Ref.~\onlinecite{OptMater.33.553}.  In both
cubic and orthorhombic phases, the extrema of the valence and
conduction bands are located at the $\Gamma$ point of the Brillouin zone (the
optical transitions are direct).

The calculations presented in this work were performed on the laboratory
computer cluster (16~cores).

\section{Conclusions}

First-principles calculations of the phonon spectra and structure of cubic
BaThO$_3$ and its distorted phases have shown that the ground-state structure
of barium thorate is $Pbnm$. The static and optical dielectric constants,
elastic moduli, heat capacity, Raman spectra, and the energy band gap in the
LDA and $GW$ approximations have been calculated for this structure. It has
been shown that the structural phase transition $Pbnm \to Imma$ can occur in
BaThO$_3$ at a temperature slightly above 300~K.


\providecommand{\BIBYu}{Yu}

\end{document}